\documentclass[preprint,preprintnumbers,amsmath,amssymb]{revtex4}

\usepackage{comment}
\usepackage{amssymb}
\usepackage{amsmath}
\usepackage[dvipsnames,usenames]{color}
\usepackage{graphicx,graphics}
\usepackage{amssymb}

\newcommand\ket[1]{\ensuremath{|#1\rangle}}
\newcommand\bra[1]{\ensuremath{\langle#1|}}
\begin{document}
\title{Entanglement distribution maximization over one-side Gaussian noisy channel}
\author{Xiang-Bin Wang}
 \email{xbwang@mail.tsinghua.edu.cn}
\author{Zong-Wen Yu}
\affiliation{Department of Physics and  the Key Laboratory of Atomic
and Nanosciences, Ministry of Education, Tsinghua University,
Beijing 100084, China}
\author{Jia-Zhong Hu}
\affiliation{Department of Physics and  the Key Laboratory of Atomic
and Nanosciences, Ministry of Education, Tsinghua University,
Beijing 100084, China}


\begin{abstract}
  The optimization of entanglement evolution for two-mode Gaussian pure states
  under one-side Gaussian
  map is studied.  Even there isn't
complete information about the one-side Gaussian noisy channel, one
can still maximize the entanglement distribution  by testing the
channel with only two specific states.
\end{abstract}

\maketitle
{\em Introduction}. The study of properties about quantum
entanglement has drawn much interest for a long
time\cite{nielsen,Vedral,Wooters,Yutin}. Although initially quantum
information processing(QIP) was studied with discrete quantum
states, it was then extended to the continuous variable (CV) quantum
states\cite{bw}. So far, many concepts and results with 2-level
quantum systems have been extended to the continuous variable case
with parallel results, such as the quantum
teleportation\cite{cvqt1}, the inseparability criteion\cite{Duan},
the degree of entanglement\cite{Giedke3,Marian}, the entanglement
purification\cite{Simon2,Plenio,Fiura}, the entanglement sudden
death\cite{csd1}, the characterization of Gaussian
maps\cite{Giedke2}, and so on. However, this does not mean {\em all}
results with 2-level quantum systems can have parallel results for
Gaussian states.

Entanglement distribution is the first step towards many novel tasks
in quantum communication and QIP\cite{nielsen}. In practice, there
is no perfect channel for entanglement distribution. Naturally, how
to maximize the entanglement after distribution is an important
question in practical QIP.  If we distribute the quantum
entanglement by sending one part of the entangled state to a remote
place through noisy channel, we can use the model of one-side noisy
channel, or one-side map.

Given the factorization law presented by Konrad et al\cite{1}, such
a maximization problem for entanglement distribution over one-side
map does not exist for the $2\times 2$ system because  any one-side
map will produce the same entanglement on the output states provided
that the entanglement of the input pure states are same. The result
has been experimentally tested\cite{sci} and also been extended
\cite{Song} recently. However, such a factorization does not hold
for the continuous variable state as shown below.  In this work, we
consider the following problem: Initially we have a bipartite
Gaussian pure state. Given a one-side Gaussian map (or a one-side
Gaussian noisy channel), how to maximize the entanglement of the
output state by taking a Gaussian unitary transformation on the
input mode before it is sent to the noisy channel. We find that by
testing the channel with only two different states, if a certain
result is verified, then we can find the right Gaussian unitary
transformation which optimizes the entanglement evolution for any
input Gaussian pure state. That is to say, we can maximize the
output entanglement even though we don't have the full information
of the one-side map. In what follows we shall first show by specific
example that the factorization law for $2\times 2$ system presented
by Konrad et al\cite{1} does not hold for Gaussian states.  We then
present an upper bound of the entanglement evolution for initial
Gaussian pure states. Based on this, we study how to optimize the
entanglement evolution over one-side Gaussian map by taking a local
Gaussian unitary transformation to the mode before sent to the noisy
channel.

{\em Output entanglement of one-side Gaussian map and single-mode
squeezing.}
Most generally, a two-mode Gaussian pure state is
\begin{equation}|g(U,V,q)\rangle=U\otimes V\ket{\chi(q)}\label{gp}\end{equation}
and $|\chi(q)\rangle=\sqrt{1-q^2}e^{qa_1^\dagger
a_2^\dagger}|00\rangle$  ($-1\le q\le 1$) is a two-mode squeezed
state (TMSS). We define map $\$$ as a Gaussian map which acts on one
mode of the state only. A Gaussian map changes a Gaussian state to a
Gaussian state only. In whatever reasonable entanglement measure,
the entanglement of a Gaussian pure state in the form of
Eq.(\ref{gp}) is uniquely determined by $q$. Therefore, we define
the {\em characteristic} value of entanglement of the Gaussian pure
state $\rho(q)=|g(U,V,q)\rangle\langle g(U,V,q)|$  as
\begin{equation}\label{Entpure}
E[\rho(q)]=|q|^2.
\end{equation}
On the other hand, any bipartite Gaussian pure state is fully
characterized by its covariance matrix (CM). Suppose the CM of state
$U\otimes V |\chi(q)\rangle$ is
\begin{eqnarray}\label{guv}
\ \Lambda =\left(\begin{array}{cc} A & C \\ C^T & B
\end{array}\right),
\end{eqnarray}
 $|q|^2$ is uniquely determined by $|A|$ (the determinant of the
matrix $A$). So, to compare the entanglement of two Gaussian pure
state, we only need to compare $|A|$ value of their covariance
matrices.

We start with the projection operator $\hat{T}_k(q_\alpha)$ which
acts on mode $k$ only:
\begin{equation}\label{operT}
\hat{T}_k(q_\alpha)=\sum^\infty_{n=0}q^n_\alpha\ket{n}\bra{n}=q_\alpha^{a_k^\dagger
a_k}.
\end{equation}
This operator has an important mathematical property
\begin{equation}\label{bchT}
\hat T_k(q_\alpha) (a_k^\dagger,a_k)\hat
T_k^{-1}(q_\alpha)=(q_\alpha a_k^\dagger, a_k/q_\alpha)
\end{equation}
which shall be used latter in this paper. For simplicity, we
sometimes omit the subscripts of states and/or operators provided
that the omission does not affect the clarity.

Define the one-mode squeezed operator $\mathcal{S}(r)= e^{r({a^\dagger}^2-a^2)}$ where $r$ is a real number and
bipartite state $|\psi_r(q_0)\rangle=I \otimes \mathcal{S}(r)
|\chi(q_0)\rangle$. We have
\\{\em Theorem 1.} Consider the
 one-side map $I\otimes \hat T(q_1)$ acting on the initial
state  $|\psi_r(q_0)\rangle$. The
 entanglement for the outcome state $I\otimes \hat T(q_1)|\psi_r(q_0)\rangle$ is a
descending function of $|r|$. Mathematically, it is to say that if
$|r_1|>|r_2|$ then
\begin{equation}\label{them1}
  E[I\otimes \hat T(q_1)|\psi_{r_1}(q_0)\rangle]
  <E[I\otimes \hat T(q_1)|\psi_{r_2}(q_0)\rangle].
\end{equation}
This theorem actually shows that there isn't a  factorization law
similar to that in $2\times 2$ states for the continuous variable
states, in whatever good entanglement measure. Using
Backer-Compbell-Horsdorff (BCH) formula, up to a normalization
factor, we have
\begin{equation}\label{tt0}
\ket{\psi_r(q_0)} =e^{-\frac{1}{2}{a_1^\dagger}^2 q_0^2 \tanh
(2r)+\frac{1}{2}{a_2^\dagger}^2\tanh(2r)+ \frac{q_0a_1^\dagger
a_2^\dagger}{\cosh(2r)}}|00\rangle.
\end{equation}
Detailed derivation of this identity is given in the appendix.
 Based on Eq.(\ref{operT}), the
one-side map $I \otimes \hat{T}(q_1)$ changes  state
$|\psi_r(q_0)\rangle$ into
\begin{equation}
|\psi'\rangle =e^{f_1{a_1^\dagger}^2+ f_2{a_2^\dagger}^2+
f_3a_1^\dagger a_2^\dagger}|00\rangle
\end{equation}
where $f_1= -\frac{1}{2}q_0^2\tanh(2r)$, $f_2= \frac{1}{2}q_1^2\tanh
(2r)$, and $f_3= \frac{q_0q_1}{\cosh(2r)}$. Here we have omitted the
normalization factor. Since we  only need the covariance matrix of
state $|\psi'\rangle$, the normalization can be disregarded because
it does not change the covariance matrix.
The characteristic function of state $\rho'=|\psi'\rangle\langle \psi'|$ has the form
\begin{equation}
C(\alpha_1,\alpha_2)= {\rm {tr}}[\rho' \hat{D}_{1}(\alpha_1)
\hat{D}_{2}(\alpha_2)]= e^{-\frac{1}{2}\bar{\alpha} \Lambda
{\bar\alpha}^T }
\end{equation}
where $\hat{D}_{k}(\alpha_{k})=e^{\alpha_{k}
a^{\dagger}_{k}-\alpha^{*}_{k} a_{k}}$ and
$\bar{\alpha}=(x_1,y_1,x_2,y_2)$ with
$\alpha_{k}=\frac{1}{\sqrt{2}}(x_k+i y_k)$.  Writing $\Lambda$ here
in the form of Eq.(\ref{guv}), we find $A={\rm diag}[b_1,b_2]$,
$C={\rm diag}[c_1,c_2]$ and $B={\rm diag}[d_1,d_2]$  with $b_1 =
-\frac{1}{2}+\frac{1+2f_{2}}{1+2f_{1}+2f_{2}+4f_{1}f_{2}-f_{3}^{2}}$,
$b_2=-\frac{1}{2}+\frac{1-2f_{2}}{1-2f_{1}-2f_{2}+4f_{1}f_{2}-f_{3}^{2}}$,
$d_1=-\frac{1}{2}+\frac{1+2f_{1}}{
 1+2f_{1}+2f_{2}+4f_{1}f_{2}-f_{3}^{2}}$,
$d_2=-\frac{1}{2}+\frac{1-2f_{1}}{1-2f_{1}-2f_{2}+4f_{1}f_{2}-f_{3}^{2}}$,
$c_1=\frac{-f_{3}}{1+2f_{1}+2f_{2}+4f_{1}f_{2}-f_{3}^{2}}$,
$c_2=\frac{f_{3}}{1-2f_{1}-2f_{2}+4f_{1}f_{2}-f_{3}^{2}}$.  The
entanglement in whatever measure of state $|\psi'\rangle$ is a
rising functional of $|A|$ and
\begin{equation}\label{xs}
|A|=
\frac{1}{4}+\frac{\scriptstyle{2q_{0}^{2}q_{1}^{2}}}{\scriptstyle{1-4q_{0}^{2}q_{1}^{2}+q_{1}^{4}+
q_{0}^{4}(1+q_{1}^{4})+(1-q_{0}^{4})(1-q_{1}^{4})\cosh{(4r)}}}.
\end{equation}
This is obviously a descending functional of $|r|$.

{\em Upper bound of entanglement evolution.}  Since $U\otimes I$ and
$I\otimes \$ $ commute, the unitary operator $U$ places no role in
the entanglement evolution under one-side map $I\otimes \$ $, and
hence we only need consider the initial state
$\ket{g(I,V,q)}=I\otimes V|\chi(q)\rangle=|\varphi(q)\rangle $. We
also define $\rho^{G}(q_\alpha)=I\otimes \$
(\ket{\varphi(q_{\alpha})}\bra{\varphi(q_{\alpha})})$.

Using Eq.(\ref{bchT}), one easily finds
$|\varphi(q=q_{a}q_{b})\rangle=\hat{T}(q_{a})\otimes I
|\varphi(q_{b})\rangle$. Since the operator $\hat
T(q_a)\otimes\uppercase\expandafter{\romannumeral1}$ and the map
$\uppercase\expandafter{\romannumeral1}\otimes \$ $ commute, there
is:
\begin{equation}\label{elecvqt}
\rho^{G}(q=q_a q_b)= \hat{T}(q_a)\otimes I \rho^{G}(q_b)\hat T^\dagger(q_a)\otimes I.
\end{equation}
 Using entanglement of
formation\cite{Marian,bennett}, we can calculate the entanglement of
the state of a Gaussian state through its optimal decomposition
form\cite{Marian}. Suppose $\rho^{G}(q_b)$ has the following optimal
decomposition\cite{Marian}:
\begin{equation}\label{rhosG}
  \rho^{G}(q_b)=U_1\otimes U_2 \rho^{s}(q_0) U^{\dagger}_{1} \otimes U^{\dagger}_{2}
\end{equation}
Here $U_1,U_2$ are two local Gaussian unitaries and $\rho^s$ is in
the form
\begin{equation}\label{decomp}
\begin{split}
\rho^s(q_0) = & \int d^2\beta_1d^2\beta_2
P(\beta_1,\beta_2) \\
& \hat
D(\beta_1,\beta_2)|\chi(q_0)\rangle\langle\chi(q_0)|\hat D^\dagger(\beta_1,\beta_2),
\end{split}
\end{equation}
where $P(\beta_1,\beta_2)$ is positive definite,
$\hat{D}(\beta_1,\beta_2)=\hat{D}_{1}(\beta_1)\otimes
\hat{D}_{2}(\beta_2)$ is a displacement operator defined as $\hat
D_k(\beta_k)=e^{\beta_ka_k^\dagger-\beta_k^* a_k}$. According to the
definition of optimal decomposition\cite{Marian,bennett}, there
don't exist any other $U_1,U_2$ and positive definite functional
$P(\beta_1,\beta_2)$ which can decompose $\rho^{G}(q_b)$ in the form
of Eq.(\ref{rhosG}) with a smaller $|q_0|$. The entanglement of
$\rho^G(q_b)$ is equal to that of a TMSS $|\chi(q_0)\rangle$, i.e.
$q_0^2$. For the Gaussian state $\rho^{G}(q_b)$ with its optimal
decomposition of Eq.(\ref{rhosG}), we define the characteristic
value of entanglement of $\rho^{G}(q_b)$ as
$E[\rho^{G}(q_b)]=|q_0|^2$.
 \\{\em Lemma
1.} For any local Gaussian unitary $U$ and operator $\hat{T}(q_a)$,
we can find $\theta,\theta'$ and $\beta''$ satisfying
\begin{equation}\label{TU}
\begin{split}
  & \hat{T}(q_a)U_1\otimes U_2 \cdot \hat{D}(\beta_1,\beta_2)\ket{\chi(q_0)}  \\
  =& \mathcal R(\theta^{\prime})\otimes \mathcal R(\theta)
  \cdot \hat{D}(\beta^{\prime}_{1},\beta^{\prime}_{2}) \cdot \hat{T}(q_{a})\mathcal S(r)\otimes
  U_2
  \ket{\chi(q_0)},
\end{split}
\end{equation}
where, $\mathcal{S}(r)$ is a squeezing operator defined earlier,
$\mathcal{R}(\theta)$ is a rotation operator defined by $\mathcal
R(\theta)(a^\dagger,a)\mathcal
R^\dagger(\theta)=(e^{-i\theta}a^\dagger,e^{i\theta}a)$,
$\beta^{\prime}_1,\beta^{\prime}_2$ and $\beta_1, \beta_2$ are
related by a certain linear transformation.
\\Proof:  Any local
Gaussian unitary operator $U_1$ can be decomposed into the product
form of $\mathcal R(\theta')\mathcal S(r) \mathcal R(\theta)$. Also,
$\mathcal{S}(r)\mathcal{R}(\theta)\otimes U_2 \cdot
\hat{D}(\beta_1,\beta_2)=
\hat{D}(\beta^{\prime\prime}_1,\beta_2'')\cdot
\mathcal{S}(r)\mathcal{R}(\theta)\otimes U_2$. Define $\hat
d=\hat{T}(q_a)\otimes I\cdot
\hat{D}(\beta^{\prime\prime}_1,\beta_2'') \cdot
\hat{T}^{-1}(q_a)\otimes I$, we have
\begin{eqnarray*}\label{nonunitary}
  & & \hat{T}(q_a)U\otimes I \cdot \hat{D}(\beta_1,\beta_2)\ket{\chi(q_0)} \nonumber \\
  &=& \hat{T}(q_a)\mathcal{R}(\theta^{\prime})\mathcal{S}(r) \mathcal{R}(\theta) \otimes I \cdot \hat{D}(\beta_1,\beta_2) \ket{\chi(q_0)} \nonumber\\
  &=& \mathcal{R}(\theta^{\prime})\otimes I \cdot \hat{d}\cdot
  \hat{T}(q_a)\mathcal{S}(r) \mathcal{R}(\theta)\otimes I \ket{\chi(q_0)} \nonumber\\
  &=& \mathcal R(\theta^{\prime})\otimes \mathcal R(\theta) \cdot \hat{D}(\beta^{\prime}_{1},\beta^{\prime}_{2}) \cdot \hat{T}(q_{a})\mathcal S(r) \otimes I \ket{\chi(q_0)}.
\end{eqnarray*}
This completes the proof of Eq.(\ref{TU}). In the second equality
above, we have used the fact $\hat T(q_a)$ and $\mathcal R(\theta')$
commute. Also, $\hat d$ there is {\em not} unitary. However, using
BCH formula and the vacuum state property $a_k|00\rangle=0$, we can
always construct a unitary operator $\hat D(\beta_1',\beta_2')$ so
that the final equality above holds. Here $\beta_1',\;\beta_2'$ are
certain linear functions of $\beta_1,\;\beta_2$.

Using  Eq.(\ref{elecvqt}) and Eq.(\ref{rhosG}) with Eq.(\ref{TU}) we
have
\begin{eqnarray}\label{Eleq}
\begin{split}
   & E[\rho^{G}(q=q_a q_b)] \\
  =& E[I\otimes U_2 \cdot \hat{T}(q_a)U_1\otimes I \rho^{s} U^{\dagger}_1\hat{T}^{\dagger}(q_a)\otimes I\cdot I\otimes U^{\dagger}_2] \\
  =& E\left[\mathcal{R}(\theta_1^{\prime})\otimes U_2 \mathcal{R}(\theta_1) \left(\int
  d^2\beta_1d^2\beta_2 P(\beta_1,\beta_2)  \right. \right. \\
  & \hat{D}(\beta^{\prime}_1,\beta^{\prime}_2) \cdot \hat{T}(q_a) \mathcal{S}(r_1)\otimes I |\chi(q_0)\rangle \langle\chi(q_0)| \mathcal{S}^{\dagger}(r_1)\hat{T}^{\dagger}(q_a) \\
  & \left. \left.  \otimes I \cdot \hat{D}^\dagger(\beta^{\prime}_1,\beta^{\prime}_2) \right) \mathcal{R}^{\dagger}(\theta_1^{\prime})\otimes \mathcal{R}^{\dagger}(\theta_1) U^{\dagger}_2 \right]  \\
  \leq& E\left[\int
  d^2\beta_1d^2\beta_2 P(\beta_1,\beta_2) \hat{D}(\beta^{\prime}_1,\beta^{\prime}_2) \cdot \hat{T}(q_a) \otimes I  \right. \\
  & \left.|\chi(q_0)\rangle \langle\chi(q_0)| \hat{T}^{\dagger}(q_a)\otimes I \cdot \hat{D}^\dagger(\beta^{\prime}_1,\beta^{\prime}_2) \right]  \\
  \leq& |q_a q_0|^2 = E[\ket{\chi(q_a)}\bra{\chi(q_a)}]\cdot E[\rho^{G}(q_{b})].
\end{split}
\end{eqnarray}
In the third step above we have used theorem 1 for the inequality
sign. This gives rise to the second theorem:
\\{\em Theorem 2.} Using the
entanglement formation as the entanglement measure, if the
entanglement of $\rho^G(q_b)$ is equal to that of TMSS $|\chi(q_0)\rangle$, the entanglement of $\rho^{G}(q=q_a q_b)$ must be not larger than that of TMSS
$|\chi(q_a q_0)\rangle$. Mathematically, it is to say that if $|q|\le |q_b|\le 1$ we have
\begin{equation}\label{q1q2}
  \frac{E[I\otimes \$(|\varphi(q)\rangle\langle \varphi(q)|)]}
  {E[I\otimes \$(|\varphi(q_b)\rangle\langle \varphi(q_b)|)]}
  \leq \frac{E[|\varphi(q)\rangle\langle \varphi(q)|]}{E[|\varphi(q_b)\rangle\langle \varphi(q_b)|]}.
\end{equation}
Here $|\varphi(q)\rangle = I\otimes V|\chi(q)\rangle$ as defined
earlier, $V$ can be any Gaussian unitary operator. Definitely, the
inequality also holds if we replace $|\varphi(q)\rangle$ by
$|g(U,V,q)\rangle$ and replace $|\varphi(q_b)\rangle$ by
$|g(U',V,q_b)\rangle$, and $U,\;U'$ can be arbitrary unitary
operators. Theorem 2 also gives rise to the following corollary.
\\{\em Corollary 1.} Given the one-side Gaussian map $I\otimes \$ $,
if the equality sign holds in formula
(\ref{q1q2}) for two specific values $q,\;q_b$ and $0<|q|<|q_b|\leq
1$, then the equality sign there holds even $q,q_b$ there are
replaced by any
  $q^{\prime},q^{\prime\prime}$, respectively, as long as $|q^{\prime}|,|q^{\prime\prime}|\in
[|q|,1]$. \\Proof. For simplicity, we first consider the case where
$q$ is replaced by any $q'$. (1) suppose $|q'|\in [|q|,|q_b|]$. The
left side of formula (\ref{q1q2}) is equivalent to $w'\cdot z'$, and
$w' = \frac{E[I\otimes \$(|\varphi(q)\rangle\langle \varphi(q)|)]}
  {E[I\otimes \$(|\varphi(q')\rangle\langle \varphi(q')|)]}$ and
  $z'
  =\frac{E[I\otimes \$(|\varphi(q')\rangle\langle \varphi(q')|)]}
  {E[I\otimes \$(|\varphi(q_b)\rangle\langle \varphi(q_b)|)]}$.
  The right side of formula (\ref{q1q2}) is equivalent to $w\cdot
z$ and $w= \frac{E[|\varphi(q)\rangle\langle
\varphi(q)|]}{E[|\varphi(q')\rangle\langle \varphi(q')|]}$ and $z=
\frac{E[|\varphi(q')\rangle\langle
\varphi(q')|]}{E[|\varphi(q_b)\rangle\langle \varphi(q_b)|]}$.
Theorem 2 itself says that $w'\le w$ and $z'\le z$. If the equality
sign holds in formula (\ref{q1q2}), we have $w'\cdot z'=w\cdot z$
hence we must have $w=w'$ and $z=z'$ which is just corollary 1 in
the case $q$ is replaced by $q'$. (2) Suppose $|q'|>|q_b|$. As we
have already known, $\rho^G(q)=\hat T(q_a)\otimes I \rho^G(q_b)$.
Consider  Eq.(\ref{TU}). Unitary $U_1$ in the optimal decomposition
of Eq.(\ref{rhosG}) must be a rotation operator only, i.e., it
contains no squeezing, for, otherwise, according to theorem 1,
$E(\rho^G(q'))$ is strictly less than $q_0^2q_a^2$ which means the
equality in formula (\ref{q1q2}) does not hold.

We denote $q'=q_b/q_c$ and $|q_c|<1$.
  We have
\begin{eqnarray}\label{sqgeq}
  & &\rho^{G}(q^{\prime}=q_b/q_c) \nonumber \\
  &=& \hat{T}^{-1}(q_c)\otimes I \rho^{G}(q_b) \left(\hat{T}^{-1}(q_c)\otimes I\right)^{\dagger} \nonumber \\
  &=& \hat{T}^{-1}(q_c)\otimes I \cdot \mathcal{R}_{1}\otimes U_2 \rho^{s} \mathcal{R}_{1}^{\dagger}\otimes U_{2}^{\dagger} \cdot \hat{T}^{-1}(q_c)\otimes I \nonumber \\
  &=& \mathcal{R}_1\otimes U_2 \cdot \int
  d^2\beta_1 d^2 \beta_2 P(\beta_1,\beta_2) \hat{D}(\beta^{\prime}_1,\beta^{\prime}_2) \nonumber \\
  & & \ket{\chi(q_0/q_c)}\bra{\chi(q_0/q_c)} \hat{D}^\dagger(\beta^{\prime}_1,\beta^{\prime}_2) \cdot \mathcal{R}_1^{\dagger}\otimes  U_2^{\dagger}.
\end{eqnarray}
Here we have used $\hat{T}^{-1}(q_c)\otimes
I\ket{\chi(q_b=q^{\prime}q_c)}= \ket{\chi(q^{\prime})}$. We have
used the optimal decomposition for $\rho^G(q_b)$ in the second
equality, and lemma 1 in the last equality above. Eq.(\ref{sqgeq})
is one possible decomposition of the state $\rho^G(q')$, but not
necessarily the optimized decomposition. Therefore,
$E[\rho^{G}(q^{\prime}=q_b/q_c)]\leq {|q_0|^2}/{|q_c|^2} =
{|q^{\prime}|^2}/{|q_b|^2}\cdot E[\rho^{G}(q_b)]$. On the other
hand, according to theorem 2, we further obtain that
$E[\rho^{G}(q_b=q^{\prime} q_c)]\leq {|q_b|^2}/{|q^{\prime}|^2}\cdot
E[\rho^{G}(q^{\prime})]$. Remark: Since here $|q'|\ge q_b$,  sign
$\le$ should be replaced by sign $\ge$ in formula (\ref{q1q2}), when
$q$ is replaced by $q'$. These two inequalities and  result of (1)
lead to
\begin{equation}\label{Eqgeq}
  \frac{E[\rho^{G}(q^{\prime})]}{E[\rho^{G}(q_b)]}= \frac{E[\ket{\chi(q^{\prime})}\bra{\chi(q^{\prime})}]}
  {E[\ket{\chi(q_b)}\bra{\chi(q_b)}]}.
\end{equation}
for any $q'$ provided that $|q|\le|q'|\le 1$. Replacing symbol $q'$
above by symbol $q''$, we have another equation. Comparing these two
equations we conclude corollary 1.
\\{\em Lemma 2}: Given any Gaussian unitaries $U,\;V$, we have
\begin{eqnarray}\label{maxent}
 E[I\otimes \$
  (U\otimes V \ket{\phi^{+}}\bra{\phi^{+}} U^{\dagger}\otimes V^{\dagger})]
  =E[I\otimes \$
  ( \ket{\phi^{+}})].
\end{eqnarray}
Here $|\phi^+\rangle$ is the maximally entangled state
 defined as the simultaneous eigenstate of position
difference $\hat{x}_1-\hat{x}_2$ and momentum sum
$\hat{p}_1+\hat{p}_2$, with both eigenvalues being 0. Also, when
$q=1$, the state $\ket{\chi(q)}=\ket{\phi^{+}}$. We shall use the
following fact.
\\{\em Fact 1:} For any local Gaussian unitary operators $U$ and $V$,
we can always find another Gaussian unitary operator $\mathcal{V}$
so that
\begin{equation}\label{maxstate}
  U\otimes V \ket{\phi^{+}}= \mathcal{V}\otimes I\ket{\phi^{+}}.
\end{equation}
Proof:  Any local Gaussian unitary operator can be decomposed into
the product form of $\mathcal R(\theta')\mathcal S(r) \mathcal
R(\theta)$. For any TMSS $|\chi(q)\rangle$ we have $\mathcal
R(\theta_1)\otimes \mathcal
R(\theta_2)|\chi(q)\rangle=\uppercase\expandafter{\romannumeral1}\otimes
\mathcal{R}(\theta_1+\theta_2)|\chi(q)\rangle$. For the maximally
TMSS $|\phi^+\rangle$ we have $\mathcal
S(r)\otimes \mathcal S(r)|\phi^+\rangle=|\phi^+\rangle$, for, the
both sides are the simultaneous eigenstates of position difference
and momentum sum, with both eigenvalues being 0. This also means
$\mathcal S(r)\otimes
\uppercase\expandafter{\romannumeral1}|\phi^+\rangle
=\uppercase\expandafter{\romannumeral1}\otimes \mathcal
S^\dagger(r)|\phi^+\rangle$. Suppose  $V=\mathcal
R(\theta_B')\mathcal S(r_B)\mathcal R(\theta_B)$, then
\begin{eqnarray}
U\otimes V \ket{\phi^{+}}= \mathcal{V}\otimes I |\phi^+\rangle
\end{eqnarray}
where $\mathcal {V}=U\mathcal{R}(\theta_B)\mathcal S^\dagger
(r_B)\mathcal R(\theta_B')$. This completes the proof of
Eq.(\ref{maxstate}). If the equality sign in formula (\ref{q1q2})
holds, we can apply  corollary 1 of  theorem 2 through replacing
$q_b$ by 1 and we obtain that $E[\rho^{G}(q^{\prime})] = |q'|^2\cdot
E[I\otimes \$ (\ket{\phi^{+}})]$. On the other hand, by using
theorem 2 and lemma 2 we have $E[\rho^{G}(q^{\prime})]\leq
|q'|^2\cdot E[I\otimes \$ (\ket{\phi^{+}})]$. This means
\begin{equation}
E[\rho^{G}(q^{\prime})]=
\max_{\{V'\}}\{ E[I\otimes
\$(\ket{g(I,V',q^{\prime})})]\}
\end{equation}
where $\rho^{G}(q')=I\otimes \$ (\ket{g(I,V,q')}\bra{g(I,V,q')})$ as
defined earlier, $\{V'\}$ is the set containing all single-mode
Gaussian unitary transformations. The  equality holds for {\em any}
$q'$  provided that the equality of formula(\ref{q1q2}) holds for
two specific values $q,\;q_b$ and $|q'|\ge |q|$. We arrive at the
following major conclusion of this Letter:\\ {\em Major conclusion}:
Suppose that we have a TMSS $|\chi(q')\rangle$. We want to maximize
the entanglement distribution over a one-side Gaussian map $I\otimes
\$ $ by taking local Gaussian unitary operation $I\otimes V'$ before
entanglement distribution. Although we don't have complete
information of the map $I\otimes \$ $, it's still possible for us to
find out a specific Gaussian unitary operation $V$ so that the
entanglement distribution is maximized over all $V'$, for an initial
state $\ket{\chi(q')}$ with {\em any} $|q'|\ge |q|$, as long as we
can find two specific values $|q_b|>|q|$, such that the equality
sign in formula (\ref{q1q2}) holds.
 Obviously, the conclusion is also correct for any initial state
which is a Gaussian pure state.

The conclusion actually says that, in verifying that $V$ can
maximize the entanglement distribution for all initial states
$\{|\chi(q')\rangle|  |q'| \ge |q|\}$, we only need to verify the
equality sign of formula (\ref{q1q2})  for two specific values.

{\em Experimental proposal}. To experimentally test our major
conclusion, we can consider the following beamsplitter channel:
Initially, beams 1 and 2 are in a TMSS,
which is the initial bipartite Gaussian pure state. Beam 3 is in a
squeezed thermal state
$\rho_{3}=\tilde{S}(u_3)\rho_{th}\tilde{S}^\dagger(u_3)$ here $\tilde{S}(u)$ is a squeezing operator defined by $\tilde{S}(u)(\hat{x},\hat{p})\tilde{S}^{\dagger}(u)=(u\hat{x},\hat{p}/u)$ and $\rho_{th}$ is a thermal state whose CM is ${\rm diag} [b_3,b_3]$.
Beam 3 together with the beamsplitter makes the one-side Gaussian
channel. A beamsplitter will transform $\hat x_2, \hat x_3$ by $
U_B(\hat x_2,\hat x_3)U_B^{-1}\longrightarrow (\hat x_2,\hat
x_3)\left(
\begin{array}{cc}
\cos\theta & \sin \theta \\ -\sin\theta & \cos\theta
\end{array}\right).
$ In an experiment, we can take, e.g., $q=0.02$ and $q_b=0.5$,
testing with many different $V$ we should find that the equality
sign in formula (\ref{q1q2}) can hold with $V=\tilde{S}(u_2=u_3)$ .
Our major conclusion is verified if we can find that the same
$V=\tilde{S}(u_3)$ always maximizes the output entanglement for any
input state $|\chi(q')\rangle$ provided that $|q'|\ge 0.02$.
Numerical calculation is shown in the following figure.
\begin{figure}[h]
  \begin{center}
  \includegraphics[width=70mm]{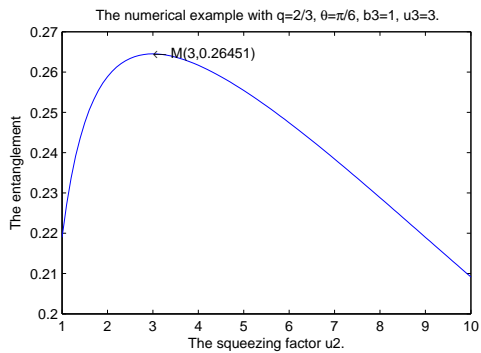}
  \caption{The entanglement with different squeezing factor $u_2$.
  The maximum entanglement obtained when $u_2=u_3=3$.
  Here we set $u_3=3$ and $q'=2/3, \theta=\pi/6, b_3=1$.}
  \end{center}
  \label{fig:fig1}
\end{figure}

In summary, we present an upper bound of the entanglement evolution
of a 2-mode Gaussian pure state under one-side Gaussian map. We show
that one can maximize the entanglement distribution over an unknown
one-side Gaussian noisy channel by testing the channel with only two
specific states. An experimental scheme is proposed.
\\{\bf Acknowledgement.}  This work was supported in part by the
National Basic Research Program of China grant nos 2007CB907900 and
2007CB807901, NSFC grant number 60725416, and China Hi-Tech program
grant no. 2006AA01Z420.
\\{\bf Appendix.} {\em Details of the proof of Eq.(\ref{tt0}).} We will use the following lemma.
\\{\em Lemma 2.} If $\mathcal{A}$ and $\mathcal{B}$ are two noncommuting operators that satisfy the conditions
\begin{equation}
  [\mathcal{A},[\mathcal{A},\mathcal{B}]]= [\mathcal{B},[\mathcal{A},\mathcal{B}]]=0,
\end{equation}
then
\begin{equation}
  e^{\mathcal{A}+\mathcal{B}}=e^{\mathcal{A}} e^{\mathcal{B}} e^{-\frac{1}{2}[\mathcal{A},\mathcal{B}]}.
\end{equation}
This is a special case of the Baker-Hausdorff theorem of group
theory\cite{Louisell}.

The squeezing operator $S(r)=e^{r({a^{\dagger}}^{2}-a^2)}$ can be
normally ordered as\cite{Barnett}
\begin{eqnarray}
  S(r)
  &=&\frac{1}{\sqrt{\cosh(2r)}} \exp\left[\frac{{a^{\dagger}}^{2}}{2} \tanh(2r)\right] \nonumber \\
  & &\cdot \exp\left[-a^{\dagger}a(\ln(\cosh(2r)))\right] \exp\left[-\frac{1}{2}a^{2}\tanh(2r)\right].
\end{eqnarray}
We neglect the constant of normalization in all the following
calculation.
\begin{eqnarray}
  I\otimes S(r)\ket{\chi(q_0)}&=& e^{r({a_2^\dagger}^2-a_2^2)}e^{q_{0}a_1^\dagger a_2^\dagger}\ket{00}
  \nonumber \\
  &=&e^{q_{0}a_1^\dagger(a_2^\dagger \cosh(2r)-a_2 \sinh(2r))}e^{r(a_2^{\dagger2}-a_2^2)}\ket{00} \nonumber \\
  &=&e^{q_{0}a_1^\dagger(a_2^\dagger \cosh(2r)-a_2 \sinh(2r))}e^{{1 \over 2}{a_2^\dagger}^2 \tanh(2r)}\ket{00}\nonumber \\
  &=& e^{{1 \over 2}{a_2^\dagger}^2 \tanh(2r)}e^{q_{0}a_1^\dagger\{a_2^\dagger
  \cosh(2r)-[a_2+a_2^\dagger \tanh(2r)] \sinh(2r)\}}\ket{00}
  \nonumber \\
  &=&e^{{1 \over 2}{a_2^\dagger}^2 \tanh(2r)}e^{q_{0}a_1^\dagger({a_2^\dagger
  \over \cosh(2r)}-a_2 \sinh(2r))}\ket{00} \nonumber \\
  &=&e^{{1 \over 2}{a_2^\dagger}^2 \tanh(2r)}e^{q_{0} a_1^\dagger
  a_2^\dagger \over \cosh(2r)}e^{-{1 \over
  2}{a_1^\dagger}^2q_{0}^2\tanh(2r)}\ket{00} \nonumber \\
\end{eqnarray}
This is just Eq.(\ref{tt0}). In the last  equality we have used
lemma 2. This completes the proof of Eq.(\ref{tt0}).


\begin{thebibliography}{99}
\bibitem{nielsen} M.~A. Nielsen and I.~L. Chuang, \textit{Quantum
    Computation and Quantum Information}, (Cambridge
    University Press, Cambridge, 2000).
\bibitem{Vedral}
V. Vedral, M.B. Plenio, M.A. Rippin and P.L. Knight, Phys. Rev. Lett. 78, 2275 (1997).
\bibitem{Wooters}W.K. Wootters, Phys. Rev. Lett. 80, 2245 (1998).
\bibitem{Yutin}T. Yu, J.H. Eberly, Science 323, 598 (2009).
\bibitem{bw}S. Braunstein and P. van Look, Rev. Mod. Phys. 77,
513 (2005); X.B. Wang, T. Hiroshima, A. Tomita, and M. Hayashi,
Phys. Rep. 448, 1 (2007).
\bibitem{cvqt1} L. Vaidman, Phys. Rev. A 49, 1473 (1994);
 S. L. Braunstein and H. J. Kimble, Phys. Rev. Lett. 80,  869
 (1998);
H.F. Hofmann, T. Ide, T. Kobayashi, and A. Furusawa, Phys. Rev. A
62, 062304 (2000);
 A. Furusawa, J. L. Sorensen, S. L. Braunstein, C. A. Fuchs, H. J. Kimble,
and  E. S. Polzik, Science 282,706 (1998).
\bibitem{Duan}
L.M. Duan, G. Giedke, J.I. Cirac, P. Zoller, Phys. Rev. Lett. 84,
2722 (2000); R. Simon, Phys. Rev. Lett. 84, 2726 (2000); R. F.
Werner and M. M. Wolf, Phys. Rev. Lett. 86, 3658 (2001);  G. Giedke,
B. Kraus, M. Lewenstein, and J.I. Cirac, Phys. Rev. Lett. 87,
167904(2001).
\bibitem{Giedke3}G. Giedke, M. M. Wolf, O. Kr¡§uger, R. F. Werner, and J.
I. Cirac, Phys. Rev. Lett. 91, 107901 (2003).
\bibitem{Marian}
P. Marian and T.A. Marian, Phys. Rev. Lett. 101, 220403 (2008).
\bibitem{Simon2}Solomon Ivan and R. Simon, arXiv:0808.1658.
\bibitem{Plenio}J. Eisert, S. Scheel, M.B. Plenio, Phys. Rev. Lett.
89, 137903 (2002).
\bibitem{Fiura}
J. Fiurasek, Phys. Rev. Lett. 89, 137904 (2002).
\bibitem{csd1}Juan Pablo Paz and Augusto J. Roncaglia, Phys. Rev. Lett. 100, 220401
(2008); A. S. Coelho AS, F. A. S. Barbosa,  K. N. Cassemiro,
  A. S. Villar,  M. Martinelli, and P. Nussenzveig, Science, 326,  823
  (2009).
\bibitem{Giedke2}G. Giedke, J.I. Cirac, Phys. Rev. A 66, 032316
(2002).
\bibitem{1}
T Konard, F.D. Melo, M, Tiersch, C. Kansztelan, A. Aragao, and A.
Buchleitner, Nature Physics, 4, 99 (2008).
\bibitem{sci}O. Jimenez Farias, C. Lombard Latune, S.P. Walborn, L.
Davidovich, P.H. Souto Ribeiro, Science, 324, 1414, (2009).
\bibitem{Song}Chang-shui Yu, X.X. Yi, and He-shan Song, Phys. Rev. A 78, 062330
(2008); Zong-Guo Li, Shao-Ming Fei, Z.D. Wang, and W.M. Liu, Phys.
Rev. A 79, 024303 (2009).
\bibitem{bennett}C.H. Bennett, D.P. DiVincenzo, J.A. Smolin, and
W.K. Wootters, Phys. Rev. A 54, 3824 (1996).
\bibitem{Louisell}W.H. Louisell, \textit{ Quantum Statistical Properties of Radiation}, (Wiley, New York, 1973).
\bibitem{Barnett}S.M. Barnett, P.M. Radmore, \textit{Methods in Theoretical Quantum Optics}, (Oxford Science Publication, Oxford, 1997).
%

\end{thebibliography}
\end{document}